\documentclass{PoS}
\usepackage{graphicx}
\usepackage[subfigure]{graphfig}
\usepackage{amsmath}
\usepackage{epsfig}
\usepackage{epstopdf}

\def\be{\begin{equation}}
\def\ee{\end{equation}}
\def\bea{\begin{eqnarray}}
\def\eea{\end{eqnarray}}
\newcommand{\Tr}{\mathrm {Tr}}

\newcommand{\nn}{\nonumber}
\newcommand{\la}{\label}

\title{Glue Spin of the Proton
\thanks{This work is supported in part by the U.S. DOE Grant part No. DE-SC0013065. This research used resources of the Oak Ridge Leadership Computing Facility at the Oak Ridge National Laboratory, which is supported by the Office of Science of the U.S. Department of Energy under Contract No. DE-AC05-00OR22725.
}}

\ShortTitle{Glue Spin/Helicity in Nuclueon}

\author{\speaker{Yi-Bo Yang}$^{1}$, Raza Sabbir Sufian$^{1}$, Andrei Alexandru$^{2}$,  Terrence Draper$^{1}$,  \mbox{Michael J. Glatzmaier$^{1}$}, and
 Keh-Fei Liu$^{1}$
\vspace*{-0.5cm}
\begin{center}
\large{
\vspace*{0.4cm}
\includegraphics[scale=0.20]{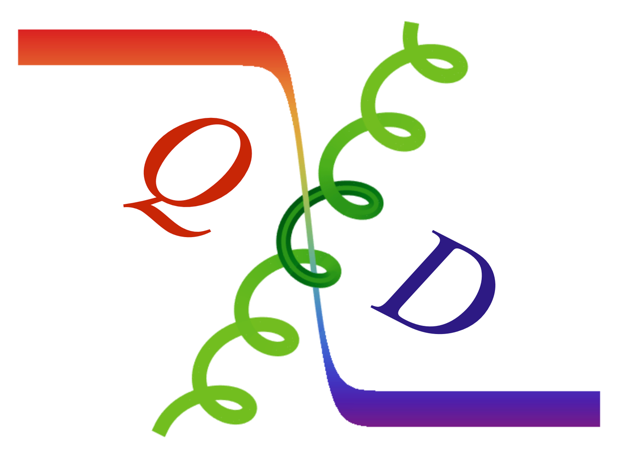}\\
\vspace*{0.4cm}
($\chi$QCD Collaboration)
}
\end{center}
\\
$^{1}$\mbox{Department of Physics and Astronomy, University of Kentucky, Lexington, KY 40506, USA}\\
$^{2}$\mbox{Department of Physics, The George Washington University, Washington, DC 20052, USA}\\
}

\abstract{We report the progress on the lattice QCD calculation of the glue spin contribution to proton spin. This calculation is carried out with valence overlap fermion on 2+1 flavor DWF gauge configurations at two lattice spacings with the momentum of the frame in the kinematic range $0\leq p^2 \leq 2$ GeV$^2$. A mild frame dependence is observed. The matching and mixing with large-momentum effective field theory are in progress. The unrenormalized result at $p^2=4$ GeV$^2$ with $O(a^2)$ correction gives $S_G$ = 0.13(3).}

\FullConference{33st International Symposium on Lattice Field Theory - LATTICE 2015\\
        July 24 - July 30, 2015\\
        Kobe International Conference Center\\
        Kobe, Japan}

\begin{document}

Deep-inelastic experiments revealed that, contrary to the naive quark model picture, the quark spin contribution to proton spin is quite small, about 25$\%$~\cite{aidala}. Recent analysis~\cite{florian,Nocera:2014gqa} of the high-statistics 2009 STAR~\cite{star} and PHENIX~\cite{phenix} {experiments at RHIC} showed evidence of non-zero glue helicity in the proton, $\Delta g$. For $Q^2=10$ GeV$^2$, the glue helicity distribution $\Delta g(x,Q^2)$ { is found to be} positive and away from zero in the momentum fraction region $0.05\leq x \leq1$ {(specifically $0.05\leq x \leq0.2$, the region in which RHIC can determine $\Delta g(x)$ much better than the other regions)}. However, the results have very large uncertainty in the region $x\le 0.05$. 

Given the importance of $\Delta g$ to address the origin of the proton spin, and the fact that large efforts are devoted to a precise experimental determination, a theoretical understanding and estimation of $\Delta g$ is highly desired. In the infinite momentum frame (IMF), the total glue helicity which is the integral of the glue helicity distribution, i.e. 
$\Delta G =\int_0^1 \Delta g(x) dx$, is defined as~\cite{manohar},
\bea \la{eq3}
\Delta G &=& \int dx \frac{i}{2xP^+}\int \frac{d\xi^-}{2\pi}e^{-ixP^+\xi^-} \langle PS | F^{+\alpha}_a (\xi^-)\mathcal{L}^{ab}(\xi^-,0)\tilde{F}^+_{\alpha,b}(0) | PS\rangle
\eea
where the light front coordinates are $\xi^{\pm}=(\xi^0\pm\xi^3)/\sqrt{2}$. The proton plane wave state is written as $|PS\rangle$, with momentum $P^\mu$ and polarization $S^\mu$. The light cone gauge-link $\mathcal{L}(\xi^-,0)$ is in the adjoint representation. It connects the gauge field tensor and its dual, $\tilde{F}^{\alpha\beta}=\frac{1}{2}\epsilon^{\alpha\beta\mu\nu}F_{\mu\nu}$, to construct a gauge invariant operator. 
After integrating the longitudinal momentum $x$, the light-cone operator for the matrix element has the following non-local expression~\cite{Hatta:2011zs,ji2},
\bea\la{eq4}
O_{\Delta G} &=& \Bigg[ \vec{E}^a(0)\times (\vec{A}^a(0)-\frac{1}{\nabla^+} (\vec{\nabla}A^{+,b})\mathcal{L}^{ba}(\xi^-,0))\Bigg]^z
\eea
In principle, one cannot evaluate this expression in the light-front frame on the lattice due to its dependence on real time, given by $\xi^-$. 

On the other hand, one can reach the glue helicity operator from the glue spin operator defined as 
\bea\la{ExA}
\vec{S}_g =  \int d^3x\ \vec{E}^a\times\vec{A}^a_{phys}
\eea
where $A_{\mu}^{phys}$ comes from the decomposition scheme as proposed in~\cite{chen1, chen2}, 
\bea
A_{\mu} = A_{\mu}^{phys} + A_{\mu}^{pure}. 
\eea

In such a scheme, $A^{phys}$ and $A^{pure}$ transform homogeneously and inhomogeneously, with respect to gauge transformation,
\bea \label{eq5}
A_\mu^{phys}  &&  \rightarrow A_\mu^{'phys}= g(x) A_\mu^{phys}g^{-1}(x) \nn \\
A_\mu^{pure} &&  \rightarrow A_\mu^{'pure} = g(x) A_\mu^{pure}g^{-1}(x)+\frac{i}{g_0}g(x)\partial_\mu g^{-1}(x), 
\eea
where $g$ is the gauge transformation matrix and $g_0$ is the coupling constant. In order to have a unique solution, conditions are set as follows: the pure gauge part does not give rise to a field tensor by itself
\bea
F_{\mu\nu}^{pure} &&=\partial_\mu A_\nu^{pure} -\partial_\nu A_\mu^{pure}+ig_0[A_\mu^{pure},\, 
A_\nu^{pure}]=0
\eea and $A_\mu^{phys}$ satisfies the non-Abelian transverse condition,
\bea \la{eq6}
 D_i\, A_i^{phys} &&= \partial_i\, A_i^{phys}-ig_0[A_i,\,A_i^{phys}] =0. 
\eea

It is shown in~\cite{ji2} that when boosting the glue spin operator $\vec{S}_g$ in Eq.~(\ref{ExA}) to IMF, the condition Eq.~(\ref{eq6}) corresponds to the light-cone gauge fixing condition $A_+^{phys}=0$ and the forward matrix element of the longitudinal glue spin operator corresponds to the glue helicity, $\Delta G$. 

In contrast to the definition of the glue helicity operator in the IMF, the matrix element of the glue spin operator in the finite momentum frame is calculable in lattice QCD. Solutions for $A_\mu^{phys}$ and $A_\mu^{pure}$ satisfying Eqs.~(\ref{eq5}-\ref{eq6}) can be obtained through a gauge link fixed in the Coulomb gauge under a gauge transformation $g_C(x)$,
\bea
U_\mu(x) = g_C(x)U^C_\mu(x)g_C^{-1}(x+a\hat{\mu})
\eea
where $U^C(x)$ is fixed in the Coulomb gauge. One can confirm that the solution for $A_\mu^{phys}$ satisfing the gauge transformation law in Eq.~(\ref{eq5}) can be obtained as in Ref.~\cite{yibo}
\bea
A_\mu^{phys} &&\equiv \frac{i}{ag_0} (U_\mu(x)-U_\mu^{pure}(x)) \nn \\
&& = g_C(x)A_\mu^C(x)g_{C}^{-1}(x) +\mathcal{O} (a),
\eea
where $U_\mu^{pure} (x)= g_C(x)g_{C}^{-1}(x+ a\hat{\mu})$.
Then the glue spin operator $\vec{S}_G$ reads,
\bea
\vec{S}_g = \int d^3x\ \Tr (g_C\vec{E}g_{C}^{-1}\times \vec{A}^C) = \int d^3x\ \Tr (\vec{E}^C\times \vec{A}^C)
\eea
where the trace $\Tr$ is taken over color indices and $\vec{E}^C$ is the electric field in the Coulomb gauge.

In order to extract the glue spin contributions to the nucleon, we compute the ratio of the disconnected three-point function to the two-point function with the source and sink of the nucleon propagator located at $t_0$ and $t$, respectively. 
The glue spin operator is inserted at the time slice $t'$ between $t_0$ and $t$. 
Then the ratio of the disconnected insertion three-point function to two-point function in the moving frame along the $z$ direction is,
\bea
R(t,t',t_0) = \frac{\langle 0|\Gamma^m_{3}\int d^3  ye^{-ip_3y_3}\chi(\vec{y},t)S_g^{3}(tÔ)\bar{\chi}(\vec{0},t_0)|0 \rangle}{\langle 0|\Gamma^e\int d^3 ye^{-ip_3y_3}\chi(\vec{y},t)\bar{\chi}(\vec{0},t_0)|0 \rangle} 
\eea
where $\chi$ is the nucleon interpolation field and $\Gamma^m_{3}/\Gamma^e$ is the polarized/unpolarized projection matrix of the proton respectively. The signal to noise ratio (SNR) can be improved by using the summed ratio method \cite{sum} and we denote the summed ratio by $SR(t,t_0) = \sum_{t'=t_0+1}^{t-1} R(t,t',t_0)$. $SR(t,t_0)$ has a linear behavior in the region where $t$ is large enough that excited-state contamination is negligible,
\bea    \label{eq:ratio_DI}
SR(t-t_0)\equiv SR(t,t_0)\ _{\overrightarrow{t  \gg 1}}\ C_0+(t-t_0) S_G+{\cal O}(e^{-\Delta E(t-t_0)}),
\eea
where $S_G $
 is the matrix element of the longitudinal glue spin operator in the proton.
 
 A preliminary attempt ~\cite{sabbir} to calculate $S_G$ on the lattice following the above prescription has been carried out on $2+1$ flavor dynamical domain-wall configurations on a $24^3\times 64$ lattice (24I) with the sea pion mass at $330$ MeV. In this proceeding, we improve the statistics on the ensemble mentioned above and carry out the calculation on another ensemble with smaller lattice spacing, i.e. the $32^3\times 64$ lattice, to check the ${\cal O}(a^2)$ correction to the glue spin. The parameters of the ensembles used in this proceeding are listed in Table~\ref{table:r0}.

To obtain $S_G$ in a relatively large momentum frame, we calculate $S_G$ for five different lattice momenta $(p= 0, 1, \sqrt{2}, \sqrt{3}, 2)$ on the 24I ensemble and seven different lattice momenta $(p= 0, 1, \sqrt{2}, \sqrt{3},  2, \sqrt{5}, \sqrt{6})$ on the 32I ensemble. All the momenta are  in unit of $2\pi/L$.  On each ensemble, five quark masses with corresponding pion mass between 250 MeV and 400MeV are used and a linear chiral extrapolation is applied to obtain the quantities at the physical point. 

\begin{table}
\begin{center}
\caption{\label{table:r0} The parameters for the RBC/UKQCD configurations. $m_s^{(s)}$ and $m_{\pi}^{(s)}$ are the strange quark mass and the pion mass of the light sea quarks on the 2+1 flavor configurations, and $N_{cfg}$ is the number of configurations used in the simulation.}
\begin{tabular}{c|ccccc}
 Symbol & $L^3\times T$  &a (fm) & $m_s^{(s)}$ (MeV) &   $m_{\pi}^{(s)}$ (MeV)  & $N_{cfg} $ \\
\hline
24I &$24^3\times 64$& 0.112(3) &120   &330  & 203    \\
32I &$32^3\times 64$& 0.084(2) & 110  &300 & 305
\end{tabular}
\end{center}
\end{table}

\begin{figure}[htb]
\centering
\includegraphics[scale=1.0]{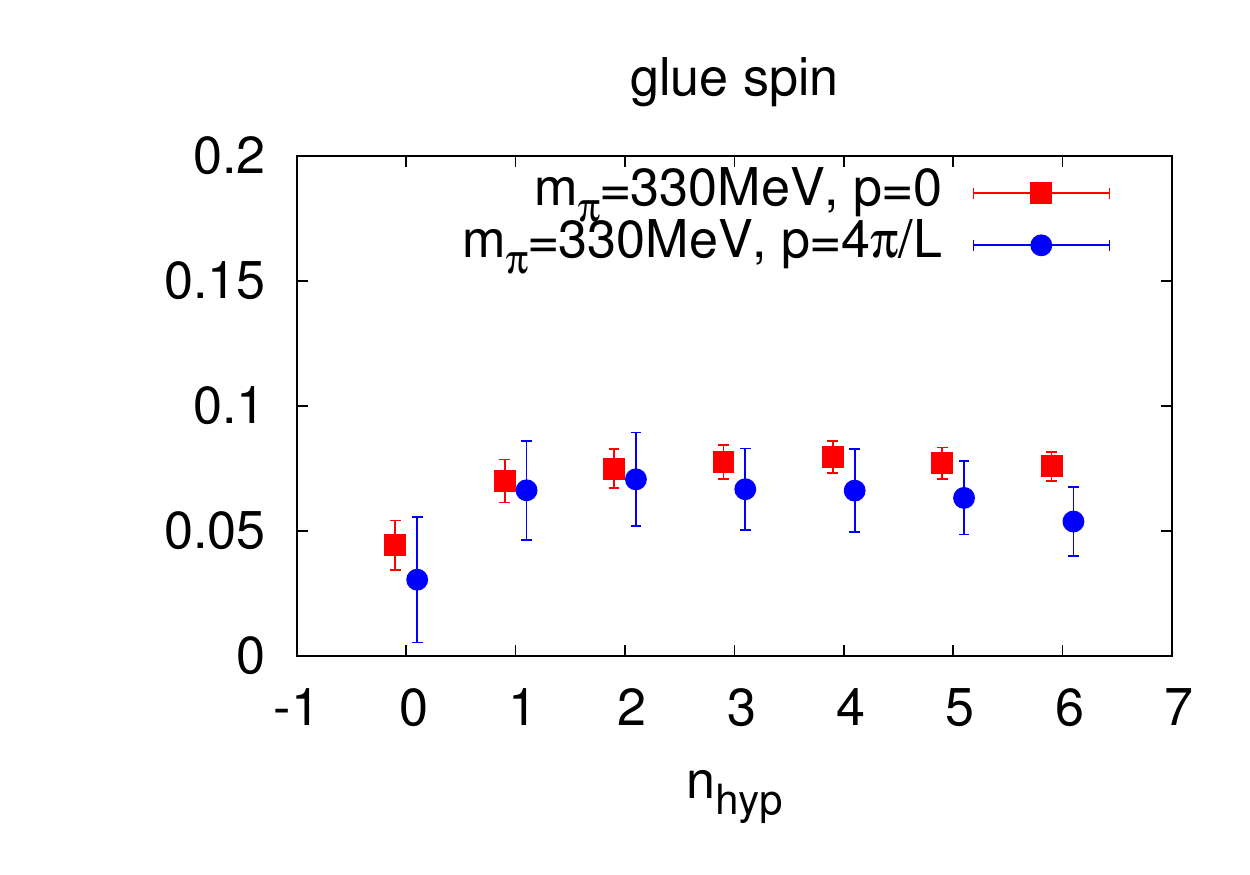}
\vspace*{-0.5cm}
 \caption{\small The HYP smearing dependence of the glue spin $S_G$, for $p=0$ (red squares) and $p=4\pi/L$ (blue dots). Both cases converge after one HYP smearing step.}
\label{fig:dep}
\end{figure}

To increase SNR, we generated the nucleon two point correlators with the source located on all the time slices on two ensembles and calculated the electric field with the clover definition,
\bea
F_{\mu\nu}^C &=& \frac{i}{8a^2g_0} (\mathcal{P}_{\mu,\nu}-\mathcal{P}_{\nu,\mu}+\mathcal{P}_{\nu,-\mu}-\mathcal{P}_{-\mu,\nu} \nn \\
&& + \mathcal{P}_{-\mu,-\nu} -\mathcal{P}_{-\nu,-\mu} + \mathcal{P}_{-\nu,\mu} -\mathcal{P}_{\mu,-\nu})
\eea
 with HYP smeared gauge links \cite{Hasenfratz:2001hp} fixed in the Coulomb gauge, where $\mathcal{P}_{\mu,\nu} = U^C_\mu(x)U^C_\nu(x+a\hat{\mu})U^{C\dagger}_\mu(x+a\hat{\nu})U^{C\dagger}_\nu(x)$ with $U^C_\mu(x)$ being the Coulomb gauge fixed Wilson link from $x+a\hat{\mu}$ to $x$. The Coulomb gauge fixing condition used here is enforced by requiring the spatial sum of the backward difference of the HYP smeared links to be zero,
 \bea
 \sum_{\mu=x,y,z} \Big[ U^C_{\mu}(x)-U^C_{\mu}(x-a\hat{\mu})\Big]=0
 \eea
 and the gauge fixed potential $A_C$ is defined by $\Big[\frac{U^C_\mu(x)-U^{C\dagger}_\mu(x)}{2iag_0}\Big]_{traceless}$. It has been observed that the central values of the glue spin matrix elements as a function of HYP smearing steps are unchanged after one step of smearing, as shown in Fig.~\ref{fig:dep}, while the SNR can be improved when more HYP smearing steps are applied. In this work, 6 and 11 steps HYP smearing are used for the 24I/32I ensembles respectively.

\begin{figure}[htb]
\centering
\includegraphics[scale=1.0]{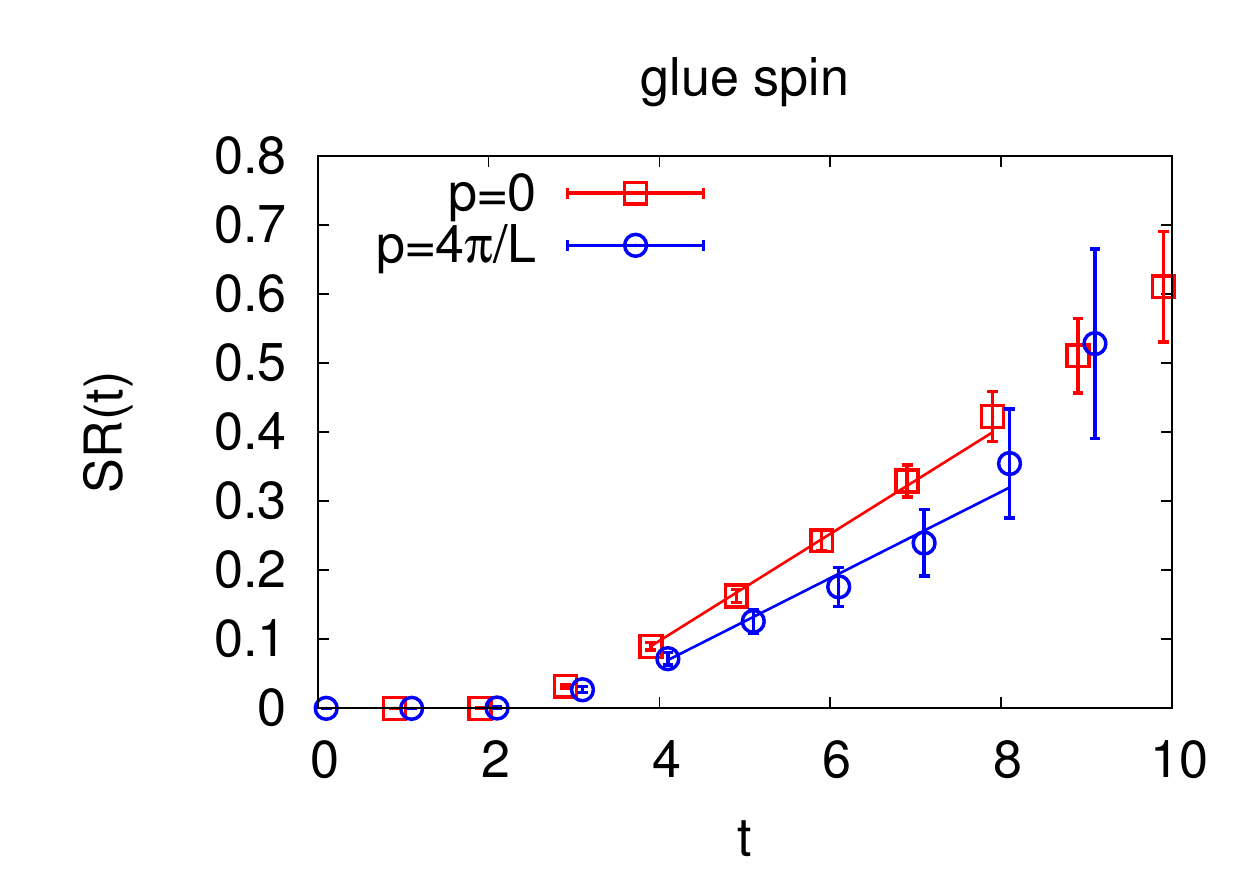}
\vspace*{-0.5cm}
 \caption{\small The summed ratio $SR(t)$ of the glue spin matrix elements for p=0 (red squares) and $p=4\pi/L$ (blue dots) cases, at the unitary point of the 24I ensemble. Linear behaviors are observed, starting from $t$=4.}
\label{fig:slope}
\end{figure}

Upon incorporating these improvements, we see a clear linear behavior of the ratio $SR(t)$ as a function of time, which indicates that excited state contamination is negligible, thus allowing a clean and reliable extraction of the matrix element. Fig.~\ref{fig:slope} shows the case of the unitary point of the 24I ensemble, as an example. 

\begin{figure}[h]
\centering
\includegraphics[scale=0.9]{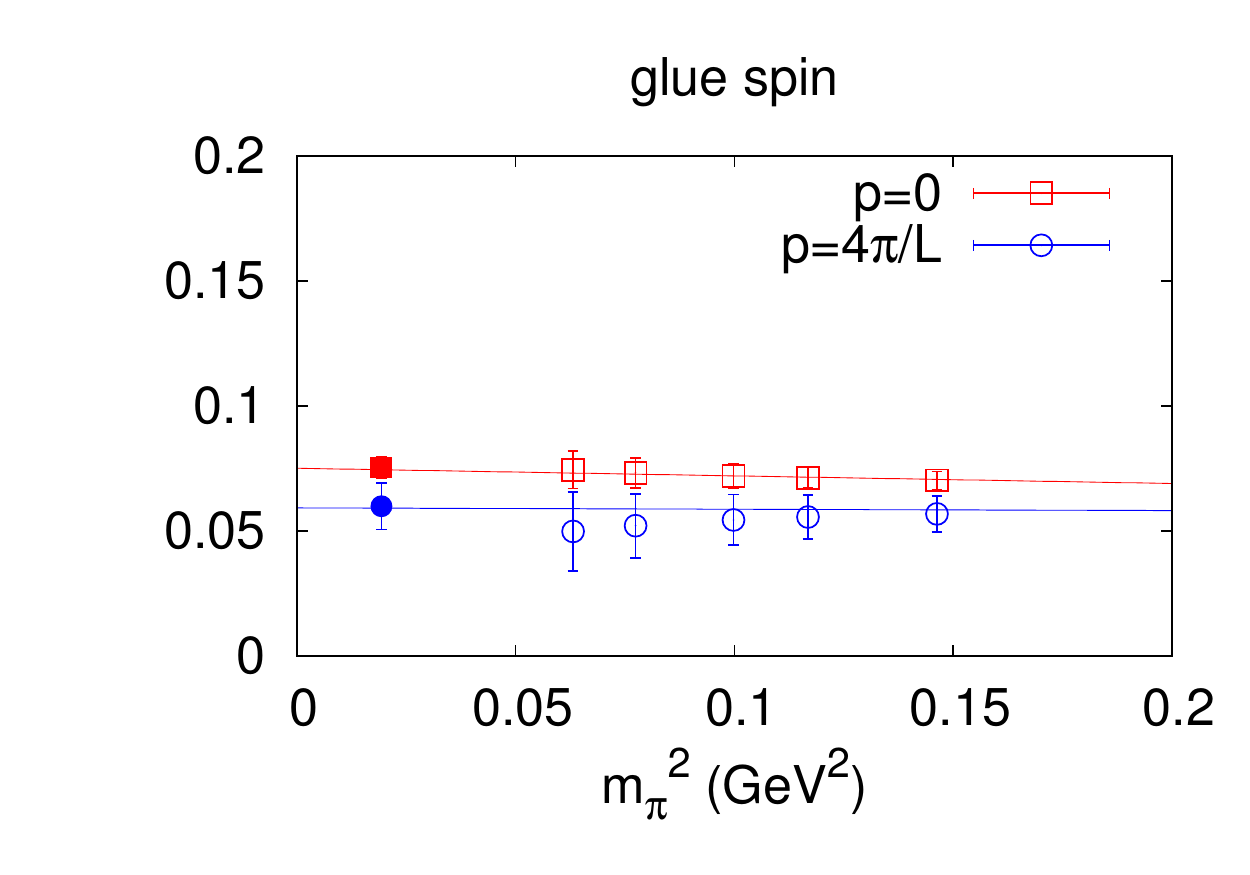}
\vspace*{-0.5cm}
 \caption{\small The quark mass dependence of the quark spin contribution in proton spin. The red squares are for zero momentum and the blue dots are for the $p=4\pi/L$ ($\sim 800$ MeV) case. These dependences are fairly mild and can be well described with a linear fit.}
\label{fig:chiral}
\end{figure}

The valence quark mass dependence is mild regardless the momentum involved, as in Fig.~\ref{fig:chiral}, which shows the case for the 24I ensemble. The sea quark mass dependence is not explored in this proceeding, and will be investigated in the future.\\

\begin{figure}[h]
\centering
\includegraphics[scale=0.9]{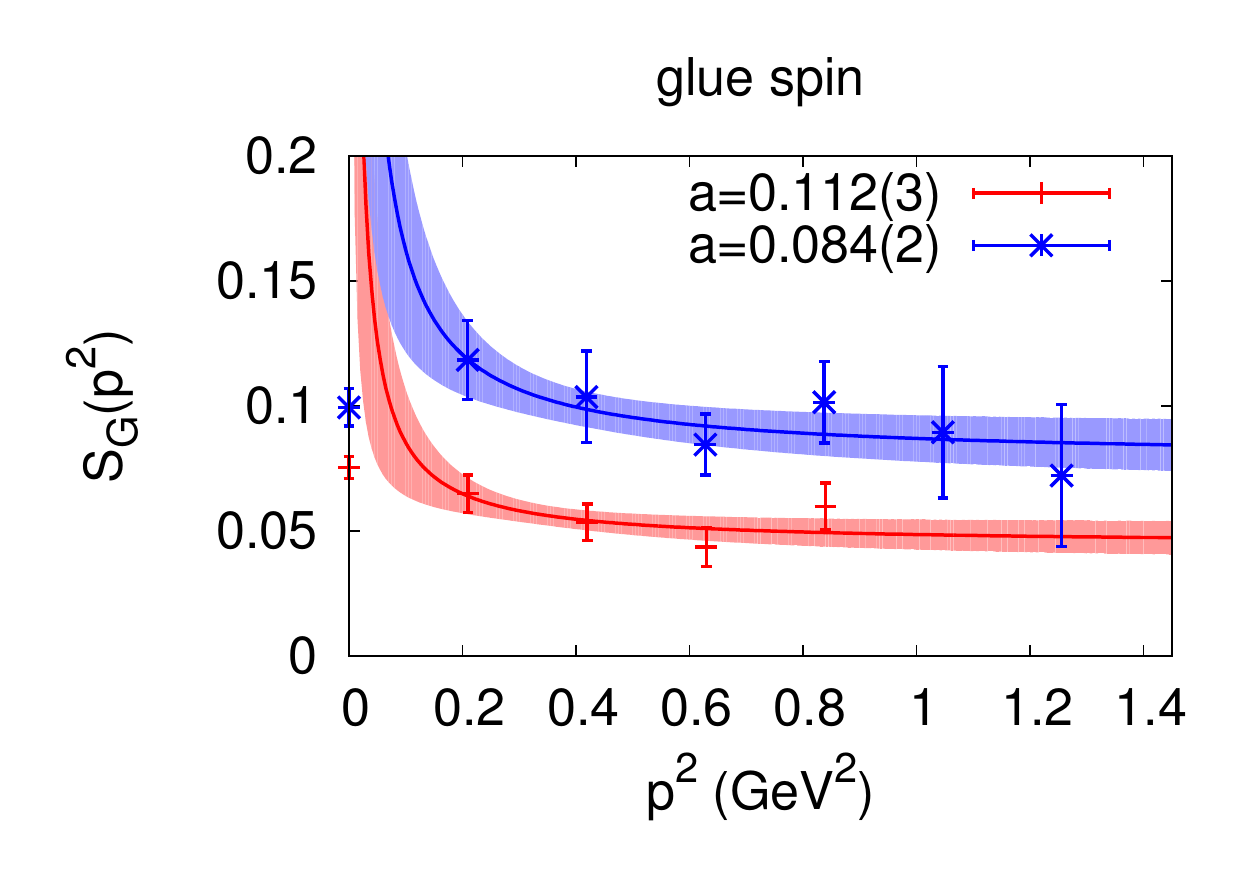}
\vspace*{-0.5cm}
 \caption{\small The result extrapolated to the physical pion mass, on two ensembles. The curves show the momentum (frame) dependence of the results.}
\label{fig:final}
\end{figure}

\begin{table}[htbp]
\begin{center}
\caption{\label{table:result} Summary of the results. The values at 2 GeV are obtained by the fit with a polynomial series of $\frac{1}{(P^z)^2}$.}
\begin{tabular}{ccc}
 a (fm)& $S_G(p=0)$ &   $S_G(p=2$ GeV)  \\
\hline
 0.112(3) & 0.075(4)  & 0.047(7)   \\
 0.084(2) &0.100(7)  & 0.084(10) \\
\end{tabular}
\end{center}
\end{table}

The glue helicity in proton spin, $\Delta G$, corresponds to the glue longitudinal spin component $S_G$ in the infinite momentum frame. An equivalent approach \cite{Ji:2014lra} is matching the glue spin under a regularization in a finite momentum frame (such as the lattice regularization) at a relatively large scale (e.g. 2 GeV or larger) to the corresponding matrix element regularized in IMF, with the large-momentum effective field theory,
\bea\label{extrapolation}
\tilde{\cal O}(P^z)=Z(P^z,1/a){\cal O}(1/a) +\frac{c_2}{(P^z)^2}+\frac{c_3}{(P^z)^4}+...
\eea
and including its mixing from the quark spin. This part of the calculation is in progress.

Setting $Z(P^z,1/a)=1$ for now,  $S_G$ can be extrapolated to $P_z$=2 GeV with Eq.~(\ref{extrapolation}). The extrapolated results at the physical point of the pion mass, for different momenta on two ensembles, are listed in Table~\ref{table:result} and plotted in Fig.~\ref{fig:final}. An extrapolation to continuum with a linear $O(a^2)$ correction based on 24I and 32I lattices gives the unrenormalized $S_G(p$=2 GeV) = 0.13(3), {which is consistent with $\langle \Delta g(Q^2)\rangle^{[0.05,0.2]}=$0.17(6) in} \cite{Nocera:2014gqa}{, if the $Q^2$ dependence is mild and the contribution from the other regions is negligible}.

\end{document}